\documentclass[12pt]{iopart}

\usepackage{amssymb}
\usepackage{graphicx}

\newcommand{\bq}{\begin{eqnarray}}
\newcommand{\eq}{\end{eqnarray}}
\newcommand{\bqn}{\begin{eqnarray*}}
\newcommand{\eqn}{\end{eqnarray*}}

\newcommand{\qq}{{\bf q}}

\begin{document}
\title{Non existence of a phase transition for the Penetrable Square
  Wells in one dimension} 

\author{Riccardo Fantoni}
\address{National Institute of Theoretical Physics (NITheP) and
  Institute of Theoretical Physics,  University of
  Stellenbosch, Stellenbosch 7600, South Africa} 
\ead{rfantoni27@sun.ac.za}

\date{\today}

\begin{abstract}
Penetrable Square Wells in one dimension were introduced for the
first time in 
[A. Santos {\sl et. al.}, Phys. Rev. E, \textbf{77}, 051206 (2008)]
as a paradigm for ultra-soft colloids. Using the Kastner, Schreiber,
and Schnetz theorem [M. Kastner, Rev. Mod. Phys., \textbf{80}, 167
(2008)] we give strong evidence for the absence of any phase
transition for this 
model. The argument can be generalized to a large class of model fluids
and complements the van Hove's theorem.
\end{abstract}

\pacs{05.70.Fh,64.60.-i,64.60.Bd,64.70.pv}

\maketitle
\section{Introduction}
The Penetrable Square Well (PSW) model in one dimension was first introduced
in \cite{Santos08} as a good candidate to describe star polymers in
regimes of good and moderate solvent under dilute conditions. The
issue of Ruelle' s thermodynamic stability was analyzed and the region
of the phase diagram for a well defined thermodynamic limit of the
model was identified. A detailed
analysis of its structural and thermodynamical properties where then
carried through at low temperatures \cite{Fantoni09} and high
temperatures. \cite{Fantoni10} 

The problem of assessing the existence of phase transitions for this
one dimensional model had never been answered in a definitive
way. Several attempt to find a gas-liquid phase transition were
carried through using the Gibbs Ensemble Monte Carlo (GEMC) technique
\cite{Frenkel-Smit,Panagiotopoulos87,Panagiotopoulos88,Smit89a,Smit89b}
but all gave negative results. Now it is well known that in three
dimensions the Square Well (SW) model admits for a particular choice
of the well parameters a gas-liquid transition. \cite{Liu05} As the
van Hove's theorem shows,
\cite{Hove50,Hemmer1970,Kincaid1976,Cuesta04} this disappears in 
one dimension. Nonetheless the PSW model in one 
dimension, being a non nearest neighbors fluid, is not analytically
solvable and since we have no hard core the van Hove's theorem does not
hold anymore. It is then interesting to answer the question whether a
phase transition is possible for it. We should also mention that we
also used the GEMC technique to probe for the transition in the three
dimensional PSW and we generally found that for a given well width there
is a penetrability threshold above which the gas-liquid transition
disappears. 

In the present work we use the Kastner, Schreiber, and Schnetz
(KSS) theorem \cite{Kastner08a,Kastner08b} to give strong analytic
evidence for the absence of any phase transition for this fluid
model.

The argument hinges on a theorem of Szeg\"o \cite{Szego58} on Toeplitz
matrices and can be applied to a large class of one dimensional fluid
models and complement the van Hove's theorem.

The paper is organized as follows: in Section \ref{sec:KSS} we state
the KSS theorem for the exclusion of phase transitions, in Section
\ref{sec:model} we describe the PSW model, 
in Section \ref{sec:numerics} we show numerically that the PSW model
satisfies KSS theorem, in Section \ref{sec:analytics} we show
analytically that the PSW model satisfies the KSS theorem, the
conclusive remarks are presented in Section \ref{sec:conclusions}. 
\section{The KSS theorem}
\label{sec:KSS}
The Kastner, Schreiber, and Schnetz (KSS) theorem \cite{Kastner08a,Kastner08b}
states the following. \\[1cm]

{\bf Theorem KSS:} {\sl Let $V_N: \Gamma_N\subseteq\mathbb{R}^N\rightarrow
\mathbb{R}$ be a smooth potential; an analytic mapping from the
configuration space $\Gamma_N$ onto the reals. Let us indicate with ${\cal
  H}^N(\qq)$ the Hessian of the potential. Indicating with
$\qq_c$ the critical points (or saddle points) of $V_N(\qq)$
(i.e. $\nabla_\qq V_N|_{\qq=\qq_c}=0$), with $k(\qq_c)$ their 
index (the number of negative eigenvalues of ${\cal
  H}^N(\qq_c)$). Assume that the potential is a Morse function
(i.e. the determinant of the Hessian calculated on all its critical
points is non zero). Whenever $\Gamma_N$ is noncompact, assume $V_N$
to be ``confining'',
i.e. $\lim_{\lambda\to\infty}V_N(\lambda\qq)=\infty,~~~\forall 
0\neq\qq\in\Gamma_N$. Consider the Jacobian densities, 
\bq
j_l(v)=\lim_{N\to\infty}\frac{1}{N}\ln\left[ 
\frac{\sum_{\qq_c\in Q_l([v,v+\epsilon])}J(\qq_c)}
{\sum_{\qq_c\in Q_l([v,v+\epsilon])} 1}\right]~, 
\eq
where
\bq
J(\qq_c)=\left|\det\frac{{\cal H}^N(\qq_c)}{2}\right|^{-1/2}~,
\eq
and
\bq
Q_l(v)=\left\{\qq_c | [V_N(\qq_c)/N=v]\wedge[k(\qq_c)=l(\mbox{\rm mod} 4)]
\right\}~.
\eq
Then a phase transition in the thermodynamic limit is excluded at any
potential energy in the interval $(\bar{v}-\epsilon,\bar{v}+\epsilon)$ if:
(i.) the total number of critical points is limited by $\exp(CN)$,
with $C$ a positive constant, (ii.) for all sufficiently small
$\epsilon$ the Jacobian densities are $j_l(\bar{v})<+\infty$ for
$l=0,1,2,3$.}  
\\[1cm]

Generally the number of critical points of the potential grows
exponentially with the number of degrees of freedom of the system. The
fact that the total number of critical points is limited by an
exponential is thought to be generically valid. \cite{Wales} We
then assume that for Morse potentials the first hypothesis of the
theorem is satisfied. So the key hypothesis of the theorem is the
second one, which can be reformulated as follows: {\sl for all}
sequences of critical points $\qq_c$ such that
$\lim_{N\to\infty}V_N(\qq_c)/N=\bar{v}$, we have  
\bq
\lim_{N\to\infty}|\det{\cal H}^N(\qq_c)|^{\frac{1}{N}} \neq 0~.
\eq

\section{The PSW model}
\label{sec:model}
The pair potential of the PSW model can be found as the $l\to\infty$
limit of the following continuous potential
\bq
\phi_l(r)=a[b-\tanh(l(r-1))]+c[\tanh(l(r-\lambda))+1]~,
\eq
where $a=(\epsilon_r+\epsilon_a)/2$,
$b=(\epsilon_r-\epsilon_a)/(\epsilon_r+\epsilon_a)$,
$c=\epsilon_a/2$, with $\epsilon_r$ a positive constant which represent
the degree of penetrability of the particles, $\epsilon_a$ a positive
constant representing the depth of the attractive well, and
$\lambda=1+\Delta$, with $\Delta$ the width of the attractive square
well. The Penetrable Spheres (PS) in one dimension are obtained as the
$\Delta\to 0$ limit of the PSW model. In the limit of
$\epsilon_r\to\infty$ the PSW reduces to the SW model.

The PSW model is Ruelle stable for $\epsilon_r/\epsilon_a>2(n+1)$ with
$n\leq\Delta<n+1$. \cite{Santos08,Fantoni10}

\begin{figure}[h!]
\begin{center}
\includegraphics[width=10cm]{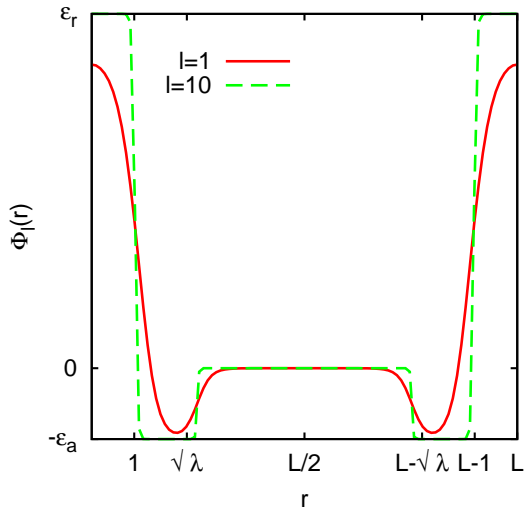}
\end{center}
\caption{Shows the potential $\Phi_l(|x|)$ for $L\gg 1$. In the plot
  we used $\epsilon_r=5,
  \epsilon_a=1, \Delta=4,$ and $L=10$, at two values of the smoothing
  parameter $l$.}
\label{fig:pot}
\end{figure}

Let us consider a pair potential of the following form 
\bq
\Phi_l(r)=\phi_l\left(2\left(\frac{L}{2\pi}\right)^2
\left[1-\cos\left(2\pi\frac{r}{L}\right)\right]\right)~.
\eq
Note that this pair potential is periodic of period $L$ and flat at
the origin, $\Phi_l^{'}(0)=0$. Moreover in the large $L$ limit
$\Phi_l(r)\approx\phi_l(r^2)$. In Fig. \ref{fig:pot} we show this
potential for different choices of the smoothing parameter $l$.

%

\section{Absence of a phase transition}
\label{sec:numerics}
In this section we will apply the KSS theorem to give numerical
evidence that there is no phase transition for the PSW model
introduced above. 

The total potential energy is
\bq
V_N(\qq)=\frac{1}{2}\sum_{i,j=1}^N\Phi_l(|x_i-x_j|)~, 
\eq 
where $\qq=(x_1,x_2,\ldots,x_N)$. If
$\lim_{N\to\infty} V_N(\qq)/N=v$ one finds $\epsilon_r/2-\epsilon_a\le
v<+\infty$. 

The saddle points $\qq_s=(x_1^s,x_2^s,\ldots,x_N^s)$ for the total
potential energy $(\nabla_\qq V_N = 0)$, can be various. We will only
consider critical point of the following kind: equally spaced
points at fixed density $\rho=N/L$, 
\bq
x_i^{\rho}=i/\rho~,~~~i=0,1,2,\ldots,N-1~.
\eq
Here we can reach
\bq
\lim_{N\to\infty}V_N(\qq_{\rho})/N=v_{\rho}~,
\eq
where for large $N$ and up to an additive constant $-\phi_l(0)/2$ we have, 
\bq
v_\rho\approx\sum_{i=0}^{N-1}\phi_l
\left(2\left(\frac{L}{2\pi}\right)^2
\left[1-\cos\left(\frac{2\pi i}{N}\right)\right]\right)~.
\eq
If $\rho \gg 1$, in the big $N$ limit we can approximate the sum by
an integral so that 
\bq \nonumber
v_\rho&\approx& \frac{N}{2\pi}\int_0^{2\pi}
\phi_l\left(2\left(\frac{L}{2\pi}\right)^2(1-\cos\alpha)\right)\,d\alpha\\
&=&\frac{N}{\pi}\int_0^2
\frac{\phi_l\left(2\left(\frac{L}{2\pi}\right)^2x\right)}
{\sqrt{1-(1-x)^2}}\, dx~,
\eq
keeping in mind that $L=N/\rho$ and $N$ is big we find in the
$l\to\infty$ limit
\bq \nonumber
v_\rho&\approx&\frac{N}{\pi}\{\epsilon_r[-\arcsin(1-z)]_{0}^{1/[2(L/2\pi)^2]}
-\epsilon_a[-\arcsin(1-z)]_{1/[2(L/2\pi)^2]}^{\lambda/[2(L/2\pi)^2]}\}\\ \label{vra}
&\approx&2\rho[\epsilon_r-\epsilon_a(\sqrt{\lambda}-1)]=v_\rho^0~,
\eq
where we used for small $z$, $\arcsin(1-z)= \pi/2-\sqrt{2z}+O[z^{3/2}]$.

For small $\rho$ in the $l\to\infty$ limit you get,
\bq
v_\rho&=&\epsilon_r/2~,~~~\rho<1/\sqrt{\lambda}\\
v_\rho&=&\epsilon_r/2-\epsilon_a~,~~~1/\sqrt{\lambda}<\rho<1
\eq

\begin{figure}[h!]
\begin{center}
\includegraphics[width=10cm]{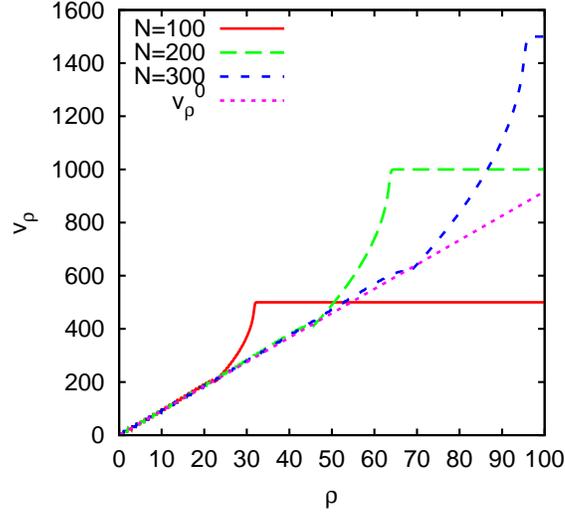}
\end{center}
\caption{Shows the behavior of $v_\rho$ as a function of the density
  $\rho$ for $N=100, 200,$ and $300$ when $\epsilon_r=5,
  \epsilon_a=1,$ and $\lambda=2$ with $l=100$. Also the theoretical
  prediction $v_\rho^0$ at big densities (Eq. (\ref{vra})) is
  shown. Notice that at fixed $N$, $v_\rho$ will saturate to $\approx
  N\epsilon_r$ for $4(L/2\pi)^2<1$ or
  $\rho>N/\pi$.} 
\label{fig:vr}
\end{figure}

For intermediate values of the density you will get a stepwise
function of the density. A graph of $v_\rho$ is shown in
Fig. \ref{fig:vr}. 

Other stationary points would be the ones obtained by dividing the
interval $L$ into $p=N/\alpha$ ($\alpha>1$) equal pieces and placing
$\alpha$ particles at each of the points $x_i^{N,p}=iL/p$,
$i=0,\ldots,p-1$. By doing so we can reach
$\lim_{N\to\infty}V_N(\qq_{N,p})/N=v_{N,p}$ where up to an additive
constant $-\phi_l(0)/2$ we have 
\bq
v_{N,p}\approx\left(\frac{N}{p}\right)\sum_{i=0}^{p-1}
\phi_l\left(2\left(\frac{L}{2\pi}\right)^2
\left[1-\cos\left(\frac{2\pi i}{p}\right)\right]\right)~.
\eq
We then immediately see that for $\rho\gg\alpha$, 
$\lim_{N\to\infty}v_{N,p}=v_\rho^0$ but for small $\rho$,
$v_{N,p}>v_\rho$.

The Hessian ${\cal H}_{i,j}^N(\qq)=\partial^2 V_N(\qq)/\partial
x_i\partial x_j$ calculated on the saddle points of the first kind can
be written as 
\bq \label{Hessianij}
{\cal H}_{i,j}^N(\qq_{\rho})&=&-\Phi_l^{''}(r_{ij})~,~~~i\neq j~,
\\ \label{Hessianii} 
{\cal H}_{i,i}^N(\qq_{\rho})&=&\sum_{j\ne i}^N\Phi_l^{''}(r_{ij})~,
\eq
where $\Phi_l^{''}(r)$ is the second derivative of $\Phi_l(r)$ and
$r_{ij}=|i-j|/\rho$.  

So the Hessian calculated on the saddle point is a circulant symmetric
matrix with one zero eigenvalue due to the fact that we have
translational symmetry $x_i^\rho=x_i^\rho\pm n/\rho$ for any $i$ and any
integer $n$. In order to break the symmetry we need to fix one point
for example the one at $x_N^\rho$. So the Hessian becomes a
$(N-1)\times(N-1)$ symmetric Toeplitz matrix (non circulant anymore)
which we call $\bar{{\cal H}}^{(N-1)}(\qq_{\rho})$.

\begin{figure}[h!]
\begin{center}
\includegraphics[width=10cm]{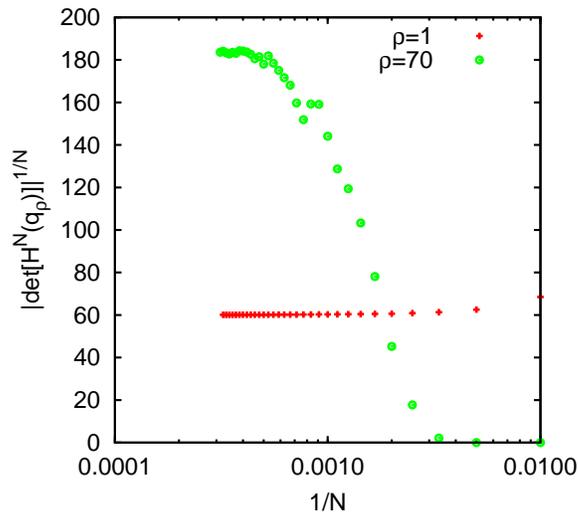}
\end{center}
\caption{Shows the behavior of $|\det \bar{{\cal H}}^N(\qq_{\rho})|^{1/N}$
as a function of $N$ at two different densities. Here we chose
$\epsilon_a=1, \epsilon_r=5, \Delta=1$, and $l=10$.} 
\label{fig:QR1}
\end{figure}

In Fig. \ref{fig:QR1} we have calculated the $|\det \bar{{\cal
  H}}^N(\qq_{\rho})|^{1/N}$ as a function of $N$ at $\rho=N/L$
fixed for $\epsilon_a=1, \epsilon_r=5, \Delta=1$, and $l=10$. One can
see that the normalized determinant of the 
Hessian does not go to zero in the large $N$ limit. So the Kastner,
Schreiber, and Schnetz (KSS) criteria \cite{Kastner08a,Kastner08b} is
not satisfied and a phase transition is excluded. The same holds
for the PS model.

\begin{figure}[h!]
\begin{center}
\includegraphics[width=10cm]{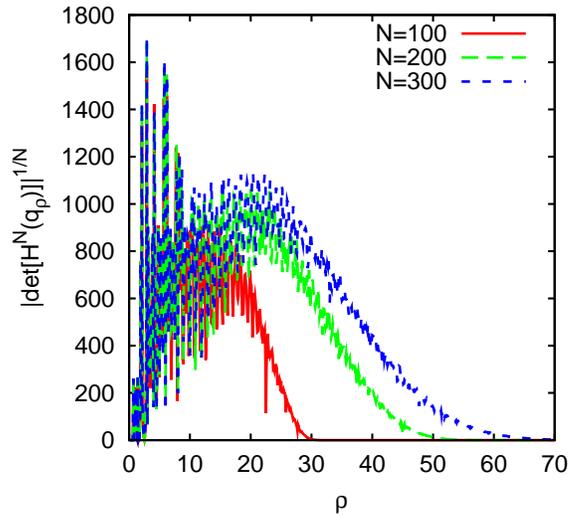}
\end{center}
\caption{Shows the behavior of $|\det \bar{{\cal H}}^N(\qq_{\rho})|^{1/N}$
as a function of $\rho$ for various $N$. Here we chose $\epsilon_a=1,
\epsilon_r=5, \Delta=1$, and $l=10$. Notice that for $\rho\lesssim
1/\sqrt{\lambda}$ then ${\cal H}^N(\qq_{\rho})\approx 0$ and also the
normalized determinant is very small. While the approach to zero at
large densities is an artifact of the finite sizes of the systems
considered.}
\label{fig:QR2}
\end{figure}

In Fig. \ref{fig:QR2} we show the dependence of $|\det \bar{{\cal
  H}}^N(\qq_{\rho})|^{1/N}$ on density for different choices of
$N$. 


\begin{figure}[h!]
\begin{center}
\includegraphics[width=10cm]{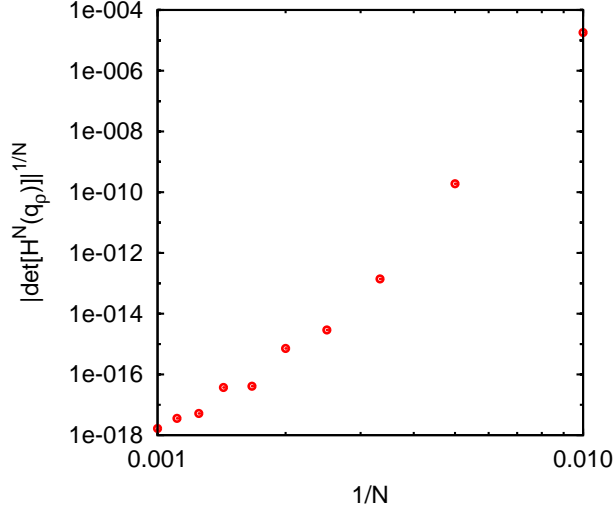}
\end{center}
\caption{Shows the behavior of $|\det \bar{{\cal H}}^N(\qq_{\rho})|^{1/N}$
as a function of $N$ for fixed $\rho=1$ in a bilogarithmic plot. Here we
chose $\epsilon=0.1$.}
\label{fig:QRSGR}
\end{figure}

A system where there is a phase transition has been proved to be the
self-gravitating ring (SGR) \cite{Nardini10} where 
$\phi_{SGR}(r)=-1/\sqrt{r+2(L/2\pi)^2\epsilon}$.
\footnote{With this choice the pair potential $\Phi_{SGR}$ would be
$2\pi\rho$ times the pair potential in the paper of Nardini and
Casetti. \cite{Nardini10}} In this case one
finds $v_\rho^0=-\rho 2\sqrt{2/\epsilon}{\cal A}(2/\epsilon)$,
with ${\cal A}(x)=\int_0^{\pi/2}d\theta\,(1+x\sin^2\theta)^{-1/2}$.
\footnote{Note that there is an error in the paper of Nardini and
Casetti. \cite{Nardini10}}
They use Hadamard upper bound to the absolute value of a determinant
to prove that indeed $\lim_{N\to\infty} |\det\bar{{\cal
  H}}^N(\qq_\rho)|^{1/N}=0$. In Fig. \ref{fig:QRSGR} we show this
numerically for a particular choice of the parameters. Actually this
result could be expected from what will be proven in the next section,
as in the large $N$ limit
for any finite $\epsilon$, $\phi_{SGR}=o(1/N)$ and $|\det\bar{{\cal
  H}}^N(\qq_\rho)|^{1/N}=o(1/N)$. This is a confirmation that theorem
KSS is not violated.  

%

\section{Limit of the normalized determinant}
\label{sec:analytics}
In this section we will give analytical evidence that there cannot be a
phase transition for the PSW model.

We need to apply to our case, Szeg\"o's theorem \cite{Szego58} for
sequences of Toeplitz matrices which deals with the behavior of the
eigenvalues as the order of the matrix goes to infinity. In particular
we will be using the following Proposition.\\[1cm]

{\bf Proposition:} {\sl Let $T_n=\{t^n_{kj}|k,j=0,1,2,\ldots,n-1\}$ be a
sequence of Toeplitz matrices with $t^n_{kj}=t^n_{k-j}$ such that
$T=\lim_{n\to\infty}T_n$ and $t_k=\lim_{n\to\infty}t^n_k$ for
$k=0,1,2\ldots$. Let us introduce  
\bq
f(x)=\sum_{k=-\infty}^\infty t_k\,e^{ikx}~,~~~x\in[0,2\pi]~.
\eq
Then there exists a sequence of Toeplitz matrices
$\tilde{T}_n=\{\tilde{t}_{kj}|k,j=0,1,2,\ldots,n-1\}$ with
$\tilde{t}_{kj}=\tilde{t}_{k-j}$ and 
\bq
\tilde{t}_k=\frac{1}{2\pi}\int_0^{2\pi}f(x)e^{-ikx}\,dx~,
\eq
such that
\bq \label{propres}
\lim_{n\to\infty}|\det T_n|^{1/n}=
\lim_{n\to\infty}|\det \tilde{T}_n|^{1/n}=
\exp\left(\frac{1}{2\pi}\int_0^{2\pi}\ln |f(x)|\,dx\right)~,
\eq
as long as the integral of $\ln |f(x)|$ exists finite.}
\\[1cm]

If the Toeplitz matrix is Hermitian then $t_{-k}=t_k^*$ and
$f$ is real valued. If moreover The Toeplitz matrix is
symmetric then $t_{-k}=t_{k}$ and additionally $f(x)=f(2\pi-x)$.

By choosing $T_N=\bar{{\cal H}}^N(\qq_{\rho})$ and
calling $t^N_{i-j}={\cal H}_{i,j}^N(\qq_{\rho})$ we
have in the $N\to\infty$ limit, with $L=N/\rho$ ($\rho$ constant),
$t_k=\lim_{N\to\infty}t^N_k$ and
\bq \nonumber
f(x)&=&\lim_{\stackrel{N\to\infty}{N \,{\rm odd}}}
\left(2\sum_{k=1}^{(N-1)/2}t^N_k\cos(kx)+t^N_0\right)\\ \label{f(x)}
&=&2\sum_{k=1}^{\infty}t_k\cos(kx)+t_0~,\\
t^N_k&=&-\Phi_l^{''}(k/\rho)~,~~~k=1,2,\ldots,(N-1)/2~,\\
t^N_0&=&-2\sum_{k=1}^{(N-1)/2} t^N_k~,
\eq
So that $f(0)=0$.
Notice that in this case the sequence of matrices $\bar{{\cal
    H}}^N(\qq_{\rho})$ does not coincide with the sequence used in
the Proposition, only the limiting matrix for large $N$
coincides. But since Szeg\"o's theorem states the limit of the
normalized determinant exists it should be independent from the
sequence chosen. An additional support to the Proposition is
presented in \ref{app:Szego}.

Now in order to prove the absence of a phase transition we need to
prove that $\int_0^{2\pi}\ln |f(x)|\,dx$ does not diverge to minus
infinity. That is we must control the way $f$ passes through zero. In
particular we do not want to have that if $x_0$ is a zero of $f$ then
\bq \label{asymptotic-f}
|f(x)|\sim e^{-1/|x-x_0|^\alpha}~,~~~x\sim x_0~,
\eq 
with $\alpha\geq 1$, which is faster than any finite power of
$(x-x_0)$. 

Now for PSW we can write $\Phi_l(r)=\Phi_l^{core}(r)+\Phi_l^{tail}(r)$.
Choose $\Phi_l^{tail}(r)=\alpha \exp(-2lr^2)$ with
$\alpha=(\epsilon_a+\epsilon_r)e^{2l}-\epsilon_ae^{2\lambda l}$.
It is then always possible to redefine the starting potential
$\Phi_l(r)$ in such a way that $\Phi_l^{core}(r)$ 
exactly vanishes for $r\ge r_{cut}>\sqrt{\lambda}$ keeping all the
derivatives at $r=r_{cut}$ continuous.
\footnote{Note that since the potential energy must be a Morse
  function (in the hypotheses of KSS theorem), we cannot take the tail
  potential $\Phi_l^{tail}(r)$ such that it exactly vanishes for $r>
  r_{cut}$. On the other hand the Gaussian decay of $\Phi_l(r)$ for
  large $r$ is sufficient to guarantee the power law behavior of $f$
  on its zeroes.}
Now in Eq. (\ref{f(x)}) for $f^{core}$ only a {\sl finite} number of
$k$ contributes to the series, namely the ones for $1\le k<\rho
r_{cut}$. So $f^{core}$ will be well behaved on its zeroes. For the
tail we get $f^{tail}(x)=-\alpha\sqrt{\pi/2l}x^2\exp(-x^2/8l)$.
 So that we will never have
$|f(x)|$ going through a zero (note that the zeroes of $f$ increase in
number as $\rho$ increases) with the asymptotically fast behavior of
Eq. (\ref{asymptotic-f}).  
This proves the absence of any phase transition for the PSW (or PS)
models. 

Note that the argument continues to hold for example for the Gaussian
Core Model (GCM) \cite{Flory1950} defined by
$\phi_{GCM}(r)=\epsilon\exp[-(r/\sigma)^2]$. In this case by choosing
$\phi(r)=\exp(-r)$ we get in the large $L$ limit $\Phi(r)=\exp(-r^2)$
and the Fourier transform of $\Phi^{''}(r)$ is
$-\sqrt{\pi}x^2\exp(-x^2/4)$ which poses no problems for the zero of
$f(x)$ at $x=0$ (note that in this case $f(x)$ is always positive for
$x>0$).   

\begin{figure}[h!]
\begin{center}
\includegraphics[width=10cm]{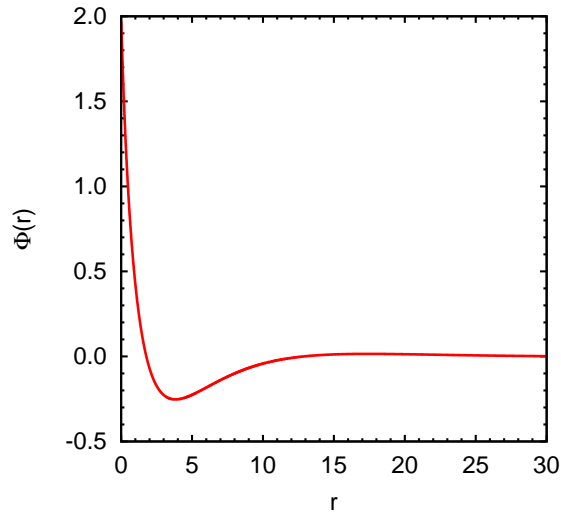}
\end{center}
\caption{Shows the pair potential $\Phi(r)=
  2[\sqrt{-ir}K_1(2\sqrt{-ir})+\sqrt{ir}K_1(2\sqrt{ir})]$ of
  the counterexample given in the text. We have $\Phi(0)=2$ and
  $\Phi(r)\propto \sin\sqrt{2r}\exp(-\sqrt{2r})$ at large $r$.} 
\label{fig:ce}
\end{figure}

The argument breaks down for example if $f(x)=-\exp(-1/|x|)$. In this case
the pair potential will be given by $\Phi(r)\sim -\int_{-\infty}^\infty
\exp(ixr)f(x)/x^2\,dx$, and one finds 
$\Phi(r)\sim 2[\sqrt{-ir}K_1(2\sqrt{-ir})+\sqrt{ir}K_1(2\sqrt{ir})]$,
where $K_n$ is the modified Bessel function of the second kind. See
Fig. \ref{fig:ce} for a plot. Also the relevant 
feature, in the pair potential, which gives the break down of the
argument for the absence of a phase transition, is the large $r$
behavior. Notice that in this case we numerically found out that the
normalized determinant tend to a finite value for large $N$. In accord
with the fact that when the hypotheses of the proposition are not
satisfied Eq. (\ref{propres}) looses its meaning. 
Considering the normalized determinant for the rescaled potential
$\Phi(r)/h(N)$, with $h(N)\to+\infty$ as $N\to\infty$, we saw that it
indeed tends to zero, indicating the presence of a phase transition.

\begin{figure}[h!]
\begin{center}
\includegraphics[width=10cm]{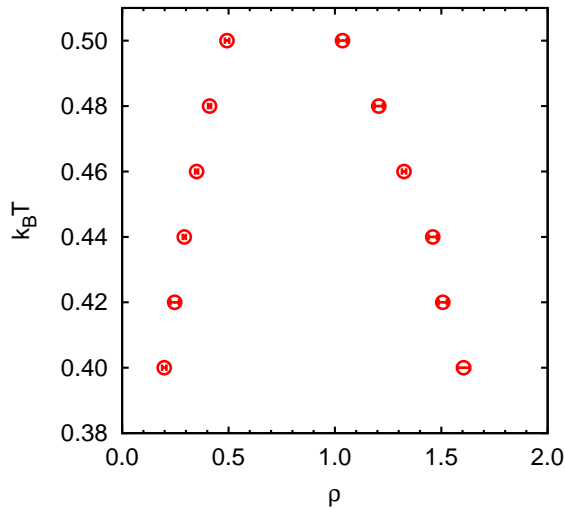}
\end{center}
\caption{Shows the gas-liquid coexistence line in the temperature
  density plane, obtained with the GEMC for $N=50$ particles \cite{chpc}
  interacting with the pair potential of Fig. \ref{fig:ce}.} 
\label{fig:pd}
\end{figure}

We simulated this model fluid and indeed we
found that it undergoes a gas-liquid phase transition. The
coexisting binodal curve is shown in Fig. \ref{fig:pd} and in Table
\ref{tab:1} we collect various properties of the two phases.
We used GEMC in which two systems can exchange both volume and
particles (the total volume $V$ and the total number of particles $N$
are fixed) in such a way to have the same pressures and chemical
potentials. We constructed the binodal for $N=50$ particles. In the
simulation we had 
$2N$ particle random displacements (with a magnitude of $0.5\sigma_i$,
where $\sigma_i$ is the dimension of the simulation box of system
$i$), $N/10$ volume changes (with a random change of 
magnitude $0.1$ in $\ln[V_1/(V-V_1)]$, where $V_1$ is the volume of one
of the two systems), and $N$ particle swap moves.
We observed that in order to obtain the binodals at different system
sizes we had to assume a scaling of the following kind: $\beta
N^\alpha=\beta_{50}50^\alpha=$constant, indicating that the model is not 
Ruelle stable (as it may be expected since it has a bounded core and a
large attractive region), and $\rho N=\rho_{50}50=$constant, where
$\beta_{50}$ and $\rho_{50}$ are the coexistence data shown in
Fig. \ref{fig:pd} and Table \ref{tab:1}. For $50\lesssim N\lesssim
100$ we found $\alpha\approx 1/2$, for $N\approx 200$ then 
$\alpha\approx 2/3$, and for $N\approx 300$ then $\alpha\approx 3/4$. 

\begin{table}[h]\scriptsize
\begin{tabular}{ccccccc}
\hline
$k_BT$ & $\rho_v$ & $\rho_l$ & $u_v$ & $u_l$ & 
$-(3\ln\Lambda)/\beta+\mu_v$ & $-(3\ln\Lambda)/\beta+\mu_l$\\
\hline
0.40	&
0.20	$\pm$0.01	&1.61	$\pm$0.03	&
-0.224  $\pm$0.009	&-0.907 $\pm$0.007	&
-0.97	$\pm$0.01	&-0.97  $\pm$0.01\\
0.42	&
0.25	$\pm$0.02	&1.51	$\pm$0.02	&
-0.26   $\pm$0.01	&-0.873 $\pm$0.008	&
-0.95	$\pm$0.01	&-0.943  $\pm$0.008\\
0.44	&
0.292	$\pm$0.007	&1.46	$\pm$0.02	&
-0.290  $\pm$0.007	&-0.854 $\pm$0.004	&
-0.938	$\pm$0.004	&-0.921  $\pm$0.006\\
0.46	&
0.350	$\pm$0.007	&1.32	$\pm$0.01	&
-0.340  $\pm$0.004	&-0.815 $\pm$0.006	&
-0.90	$\pm$0.01	&-0.89  $\pm$0.02\\
0.48	&
0.411	$\pm$0.007	&1.21	$\pm$0.02	&
-0.370  $\pm$0.003	&-0.77  $\pm$0.01	&
-0.886	$\pm$0.003	&-0.86  $\pm$0.01\\
0.50	&
0.49	$\pm$0.01	&1.04	$\pm$0.02	&
-0.420  $\pm$0.006	&-0.71  $\pm$0.01	&
-0.87	$\pm$0.01	&-0.862 $\pm$0.006\\
\hline 
\end{tabular}
\caption{Gas-liquid coexistence data ($T,\rho_i,u_i,\mu_i$ are
respectively the temperature the density, the internal energy per
particle, and the chemical potential of the vapor $i=v$ or liquid
$i=l$ phase. $\beta=1/k_BT$ and $\Lambda$ is the de Broglie thermal
wavelength.) from GEMC of $N=50$ particles \cite{chpc}.} 
\label{tab:1}
\end{table}

We then added an hard core to the potential
\bq
\Phi(r)=\left\{
\begin{array}{ll}
\epsilon & r<1\\
2[\sqrt{-ir}K_1(2\sqrt{-ir})+\sqrt{ir}K_1(2\sqrt{ir})] & r\ge 1
\end{array}
\right.~,
\eq
with $\epsilon$ a positive large number, and we saw, through GEMC,
that the corresponding fluid still admitted 
a gas-liquid phase transition (without $N$ scaling of the densities
$\rho<1$) in accord with the expectation that 
are the large $r$ tails of the potential that make this model singular
from the point of view of our argument. 

For fluids with a pair potential $\Phi$ given by a hard core and a
$-1/r^\alpha$ tail we can take the $\Phi^{''}(r)=0$ for $r<1$ and
$\Phi^{''}(r)=-\alpha(\alpha-1)/r^{\alpha-2}$ for $r>1$, and the
resulting $f$ function (the Fourier transform of $-\Phi^{''}$) is such
that $\ln|f(x)|$ has non-integrable zeros.  So this class of models
does not fall under the hypotheses pf the proposition. And it is well
known that when $1<\alpha<2$ the corresponding fluid admits a phase
transition \cite{Hemmer1970}. 
 
\section{Conclusions}
\label{sec:conclusions}

Using KSS theorem and a limit theorem of Szeg\"o on Toeplitz matrices
we were able to give strong evidence for the exclusion of phase
transitions in the phase diagram of the PSW (or PS) fluid. 
The argument makes use of the fact that the  
smoothed pair potential amongst the particles has an $r$ cutoff. Even
if we just considered two classes of stationary points, {\sl i.e.} the
equally spaced points and equally spaced clusters, we believe that our
argument give strong indications of the absence of a phase transition. 

Our argument applies equally well to model fluids with large $r$ tails
in the pair potential decaying in such a way that the condition of
Eq. (\ref{asymptotic-f}) does not hold. For example it applies to the
Gaussian Core Model. We believe this to be a rather large class of
fluid models.

We give an example of a model fluid which violates the condition
of Eq. (\ref{asymptotic-f}) and find through GEMC simulations that it
indeed has a gas-liquid phase transition.

Our argument does not require the fluid to be a nearest neighbor one, for
which it is well known that the equation of state can be calculated
analytically \cite{Salsburg53,Corti98,Heying04}. 
We think that our argument can be a good candidate to complement the well
known van Hove theorem for such systems that violates the
hypotheses of the hard core impenetrability of the particles and of
the compactness of the support of the tails. 

\appendix
\section{Alternative support to the Szeg\"o result} 
\label{app:Szego}
Our original matrix ${\cal H}^N(\qq_{\rho})$ is a circulant
matrix
\bq
{\cal H}^N(\qq_{\rho})=\left(
\begin{array}{cccccc}
h^N_0 & h^N_1 & h^N_2 & h^N_3 & \cdots & h^N_{N-1}\\
h^N_{N-1} & h^N_0 & h^N_1 & h^N_2 & \cdots & h^N_{N-2}\\
h^N_{N-2} & h^N_{N-1} & h^N_0 & h_1 & \cdots & h^N_{N-3}\\
\vdots & & & \ddots & & \vdots\\
h^N_1 & h^N_2 & h^N_3 & h^N_4 & \cdots & h^N_0
\end{array}
\right)~,
\eq
We have numerically checked that the determinant of ${\cal
  H}^N(\qq_{\rho})$ with one row and one column removed converges in
the large $N$ limit to the product of the non-zero eigenvalues of the
matrix ${\cal H}^N(\qq_{\rho})$.
\footnote{We have checked numerically that this property continues to
hold as long as the circulant matrix is a symmetric one.}

Let us assume that $N=2n+1$ is odd. Then our matrix has the following
additional structure
\bq \nonumber
h^N_{i}&=&\tilde{h}^N_{i}~,~~~i=1,\ldots,n\\
h^N_{n+i}&=&\tilde{h}^N_{n-(i-1)}~,~~~i=1,\ldots,n
\eq
The eigenvalues of $H^N$ will be given by \cite{Davis}
\bq
\psi_m=\sum_{k=0}^{N-1}h^N_k e^{-\frac{2\pi}{N}imk}~,~~~m=0,1,\ldots,N-1
\eq
with the additional constraint (see
Eqs. (\ref{Hessianij})-(\ref{Hessianii})) that 
\bq
\psi_0=\sum_{k=0}^{N-1}h^N_k=0~.
\eq
The eigenvalues can be rewritten as follows
\bq
\psi_m=\tilde{h}^N_0+\sum_{k=1}^n\tilde{h}^N_ke^{-\frac{2\pi}{N}imk}
+\sum_{k=1}^n\tilde{h}^N_{n-(k-1)}e^{-\frac{2\pi}{N}im(n+k)}~.
\eq
Introducing the summation index $j=n-k+1$ in the last sum we then
obtain
\bq \nonumber
\psi_m&=&\tilde{h}^N_0+\sum_{k=1}^n\tilde{h}^N_ke^{-\frac{2\pi}{N}imk}
+\sum_{j=n}^1\tilde{h}^N_{j}e^{+\frac{2\pi}{N}imj}\\
&=&\sum_{k=-n}^nt^N_ke^{-\frac{2\pi}{N}imk}~,
\eq
with $n=(N-1)/2$ and $t^N_k=t^N_{-k}=\tilde{h}^N_k$ for $k=1,2,\ldots,n$.

We take the logarithm of the absolute value of the product of the
non-zero eigenvalues to find
\bq
{\cal P}=\frac{1}{N}\ln\left|\prod_{m=1}^N\psi_m\right|=
\frac{1}{N}\sum_{m=1}^N\ln|\psi_m|~.
\eq
Now in the large $N$ limit we have $t_k=\lim_{N\to\infty}t^N_k$ for
$k=0,1,2,\ldots$ and
\bq
\psi_m&\sim&\sum_{k=-\infty}^\infty t_ke^{-\frac{2\pi}{N}imk}
\sim f\left(\frac{2\pi}{N}m\right)~,\\
{\cal P}&\sim&\frac{1}{N}\sum_{m=1}^N
\ln\left|f\left(\frac{2\pi}{N}m\right)\right|
\sim\frac{1}{2\pi}\int_0^{2\pi}\ln|f(x)|\,dx~,
\eq
where in the last passage we have transformed the sum into an integral.

\ack
We would like to thank Prof. Michael Kastner for his carefull guidance
in the development of the work.
Many thanks to Dr. Izak Snyman and Prof. Robert M. Gray for
helpful discussions regarding the Toeplitz matrices and Dr. Lapo
Casetti for proofreading the manuscript before publication.
\section*{References}
\bibliographystyle{unsrt}
\bibliography{1dpsw}
\end{document}